\documentstyle[12pt]{article}
\begin{document}

\vspace{.5cm}

\begin{flushright}
JINR preprint E2-97-407
\end{flushright}
\begin{center}
\Large{The system of three vortexes of two dimensional ideal hydrodynamics
as a new example of the (integrable) Nambu-Poisson mechanics}

\end{center}

\vspace{.5cm}

\begin{center}
\large{N.MAKHALDIANI\footnote{E-mail: \ mnv@lcta.jinr.dubna.su},}
\end{center}
\begin{center}
\large{Joint Institute for Nuclear Research, \\
Dubna, Moscow region, 141980, Russia}
\end{center}

\vspace{1cm}

\begin{center}
ABSTRACT
\end{center}

A Nambu-Poisson formulation of the system of three ordinary
differential equations describing dynamics of three vortexes of the
ideal two-dimensional hydrodynamics is given. The system is integrated by
quadratures.

\vspace{6cm}

\pagebreak

{\bf 1}. The equation of motion of a system of three vortexes can be
put in the form ~\cite{1,3}
\begin{eqnarray}
\dot M_{1}&=&\Gamma_{1}M_1(M_2-M_3), \\  \nonumber
\dot M_{2}&=&\Gamma_{2}M_2(M_3-M_1), \\  \nonumber
\dot M_{3}&=&\Gamma_{3}M_3(M_1-M_2).
\end{eqnarray}
Indeed, it is well known \cite{2} that the system of $N$ vortexes can be
described by the following system of differential equations:
\begin{equation}
\dot Z_n=i\sum_{m\not=n}^{N}\frac{\Gamma_m}{Z_n^*-Z_m^*}.    \label{1}
\end{equation}
Then it is easy to verify that the quantities
\begin{eqnarray}
M_1&=&|Z_2-Z_3|^2, \\ \nonumber
M_2&=&|Z_3-Z_1|^2, \\ \nonumber
M_3&=&|Z_1-Z_2|^2
\end{eqnarray}
satisfy the system (1) after changing the time parameter as follows:
\begin{equation}
dt=\frac{M_1M_2M_3}{4S_{\Delta}}d\tau=(M_1M_2M_3)^\frac{1}{2}Rd\tau,  \label{2}
\end{equation}
where $S_{\Delta}$ is the area of the triangle with vertexes in the
centres of the vortexes and $R$ is the radius of the circle with the vortexes
on it.

 The system (1) has the integrals of motion
\begin{eqnarray}
H_{1}&=&\sum_{i=1}^{3}\frac{M_{i}}{\Gamma_{i}}, \\  \nonumber
H_{2}&=&\sum_{i=1}^{3}\frac{lnM_{i}}{\Gamma_{i}}
\end{eqnarray}
\noindent
and can be presented in the form
\begin{eqnarray}
\dot M_{i}&=&\omega_{ijk}\frac{\partial H_1}{\partial M_j}\frac{\partial H_2}
{\partial M_k} \\  \nonumber
&=&\{M_{i},H_1,H_2\}=\omega_{ijk}\frac{1}{\Gamma_{j}}\frac{1}{\Gamma_{k}M_k},
\end{eqnarray}
where
\begin{eqnarray}
\omega_{ijk}&=&\epsilon_{ijk}\rho,\\  \nonumber
\rho&=&\Gamma_{1}\Gamma_{2}\Gamma_{3}M_1M_2M_3  \nonumber
\end{eqnarray}
and the Nambu-Poisson bracket of the functions $A,B,C$ on the
three-dimensional phase space $M^3$ is
\begin{eqnarray}
\{ A,B,C\}=\omega_{ijk}\frac{\partial A}{\partial M_i}\frac{\partial B}
{\partial M_j}\frac{\partial C}{\partial M_k}.
\end{eqnarray}

The fundamental bracket is
\begin{eqnarray}
\{ M_1,M_2,M_3\}=\omega_{ijk}.
\end{eqnarray}

Then we can again change the time parameter as
\begin{eqnarray}
d\tau =\rho du
\end{eqnarray}
and obtain Nambu's mechanics \cite{4}
\begin{eqnarray}
\dot M_{i}=\epsilon_{ijk}\frac{\partial H_1}{\partial M_j}
\frac{\partial H_2}{\partial M_k},    \nonumber
\end{eqnarray}
\begin{eqnarray}
\dot M_{1}&=&\frac{M_2-M_3}{\Gamma_2\Gamma_3M_2M_3}, \\ \nonumber
\dot M_{2}&=&\frac{M_3-M_1}{\Gamma_3\Gamma_1M_3M_1}, \\ \nonumber
\dot M_{3}&=&\frac{M_1-M_2}{\Gamma_1\Gamma_2M_1M_2}.
\end{eqnarray}

{\bf 2}. The second-order, ternary, Nambu-Poisson structure (6-9) reduces to
the two first-order, binary, (Nambu-)Poisson structures\cite{3}
\begin{eqnarray}
\{ M_i,M_j\}_1&=&(\{ M_i,M_j,H_1\}=\omega_{ijk}\frac{\partial H_1}{\partial M_k}
=\omega_{ijk}\frac{1}{\Gamma_{k}})=\omega^1_{ij},  \\ \nonumber
\{ M_i,M_j\}_2&=&(\{ M_i,M_j,H_2\}=\omega_{ijk}\frac{\partial H_2}{\partial M_k}
=\omega_{ijk}\frac{1}{\Gamma_{k}M_k})=\omega^2_{ij}.
\end{eqnarray}

These Poisson structures are reducible, there are nontrivial functions
$H_1$ and $H_2$ for which hold
\begin{eqnarray}
\{ A,H_1\}_1&=&0,  \\ \nonumber
\{ A,H_2\}_2&=&0
\end{eqnarray}
for any function $A$.

{\bf 3}. The variables $M_n$, $n=1,2,3,$ are non-negative (semi-bounded), so it
is convenient to replace them with free variables $x_n$
\begin{eqnarray}
x_n=~lnM_n, \ n=1,2,3.
\end{eqnarray}
The equation of motion (1), integrals (5) and Nambu-Poisson structures (8-12)
take the following form:
\begin{eqnarray}
\dot x_{1}&=&\Gamma_{1}(e^{x_2}-e^{x_3}), \\  \nonumber
\dot x_{2}&=&\Gamma_{2}(e^{x_3}-e^{x_1}), \\  \nonumber
\dot x_{3}&=&\Gamma_{3}(e^{x_1}-e^{x_2}), \\
H_{1}&=&\sum_{i=1}^{3}\frac{e^{x_i}}{\Gamma_{i}}, \\  \nonumber
H_{2}&=&\sum_{i=1}^{3}\frac{x_{i}}{\Gamma_{i}}, \\
\dot x_{i}&=&\Gamma_{1}\Gamma_{2}\Gamma_{3}\epsilon_{ijk}
\frac{\partial H_1}{\partial x_j}
\frac{\partial H_2}{\partial x_k} \\  \nonumber
&=&\{ x_{i},H_1,H_2\}=\{ x_{i},H_1\}_2=-\{ x_{i},H_2\}_1, \\
\{ A,B,C\}&=&\Gamma_{1}\Gamma_{2}\Gamma_{3}\epsilon_{ijk}
\frac{\partial A}{\partial x_i}\frac{\partial B}{\partial x_j}
\frac{\partial C}{\partial x_k},\\ \nonumber
\{ A,B\}_1&=&\omega_{ij}^1\frac{\partial A}{\partial x_i}
\frac{\partial B}{\partial x_k}, \\   \nonumber
\{ A,B\}_2&=&\omega_{ij}^2\frac{\partial A}{\partial x_i}
\frac{\partial B}{\partial x_j},\\
\omega_{ij}^1&=&\Gamma_{1}\Gamma_{2}\Gamma_{3}\epsilon_{ijk}
\frac{\partial H_1}{\partial x_k} \\   \nonumber
&=&\Gamma_{1}\Gamma_{2}\Gamma_{3}\epsilon_{ijk}\frac{1}{\Gamma_k x_k}, \\
\nonumber
\omega_{ij}^2&=&\Gamma_{1}\Gamma_{2}\Gamma_{3}\epsilon_{ijk}
\frac{\partial H_2}{\partial x_k} \\ \nonumber
&=&\Gamma_{1}\Gamma_{2}\Gamma_{3}\epsilon_{ijk}\frac{1}{\Gamma_k}.
\end{eqnarray}

{\bf 4}. For the system of three equations (15) we have two integrals of
motion (16), so the system (15) is integrable by quadratures \cite{5,6}.
>From $H_2$ it follows that
\begin{eqnarray}
x_3=\Gamma_3(H_2-\frac{x_1}{\Gamma_1}-\frac{x_2}{\Gamma_2}).
\end{eqnarray}
Inserting (20) into the expression of $H_1$, we find that
\begin{eqnarray}
\frac{\exp{x_2}}{\Gamma_2}+\frac{\exp(-\frac{\Gamma_3}{\Gamma_2}x_2)}
{\Gamma_3}\exp(\Gamma_3(H_2-\frac{x_1}{\Gamma_1}))=
H_1-\frac{\exp{x_1}}{\Gamma_1}.
\end{eqnarray}
Now we see that $x_2$ can be found from (21) as an elementary function of
$x_1$, when
\begin{eqnarray}
\Gamma_3 &=&-\Gamma_2,   \\      \nonumber
         &=&-2\Gamma_2,  \\               \nonumber
         &=&-3\Gamma_2,  \\               \nonumber
         &=&-4\Gamma_2,  \\               \nonumber
         &=&\Gamma_2,  \\               \nonumber
         &=&\frac{1}{2}\Gamma_2, \\       \nonumber
         &=&\frac{1}{3}\Gamma_2.
\end{eqnarray}
For the general case, equation (21) defines $x_2$ as a new trancendental
function $n_1(x_1)$.
Then the equation for $x_1$ takes the form
\begin{eqnarray}
\dot{x}_1=\Gamma_1(e^{n_1(x_1)}(1+\frac{\Gamma_3}{\Gamma_2})-\Gamma_3H_1)
+\Gamma_3e^{x_1}\equiv n_2(x_1)
\end{eqnarray}
and $x_1$ is defined by the following quadrature:
\begin{eqnarray}
N(x_1)\equiv\int^{x_1}_{x_{10}}\frac{dx}{n_2(x)}=\tau -\tau_0.
\end{eqnarray}

Elsewhere we consider general methods of non-linear (Nambu-)Poisson algebras
\cite{7} analysis for our model as well as detailed analysis of the formal
solution.
\vspace*{10mm}

  It is a pleasure to thank Dr.~S.A.~Gogilidze and other members of the
seminar $\odot NM\pi$ for stimulating discussion.

\end{document}